# Mapping the near-field spin angular momenta in the structured surface plasmon polaritons field

C.-C. Li, P. Shi, L.-P. Du* and X.-C. Yuan*

*Nanophotonics Research Centre, Institute of Microscale Optoelectronics, Shenzhen University, Shenzhen, 518060, China.*
*Corresponding authors: lpdu@szu.edu.cn and xcyuan@szu.edu.cn*

Optical spin angular momenta in a confined electromagnetic field exhibit remarkable difference with their free space counterparts, in particular, the optical transverse spin that is locked with the energy propagating direction lays the foundation for many intriguing physical effects such as unidirectional transportation, quantum spin Hall effect, photonic Skyrmion, etc. In order to investigate the underlying physics behind the spin-orbit interactions as well as to develop the optical spin-based applications, it is crucial to uncover the spin texture in a confined field, yet it faces challenge due to their chiral and near-field vectorial features. Here, we propose a scanning imaging technique which can map the near-field distributions of the optical spin angular momenta with an achiral dielectric nanosphere. The spin angular momentum component normal to the interface can be uncovered experimentally by employing the proposed scanning imaging technique and the three-dimensional spin vector can be reconstructed theoretically with the experimental results. The experiment is demonstrated on the example of surface plasmon polaritons excited by various vector vortex beams under a tight-focusing configuration, where the spin-orbit interaction emerges clearly. The proposed method, which can be utilized to reconstruct the photonic Skyrmion and other photonic topological structures, is straightforward and of high precision, and hence it is expected to be valuable for study of near-field spin optics and topological photonics.

## Introduction

It is well known that light carries angular momentum (AM), which plays a critical role in light–matter interactions[1–4]. The AM can be decomposed into two parts: orbital angular momentum (OAM) and spin angular momentum (SAM). The OAM is associated with the phase structure and trajectory of light[1-3]. On the contrary, the SAM is an intrinsic property of light and is associated with the helicity of polarized light[4]. The intensive investigation on SAM and their interactions with OAM yielded plentiful physical phenomena, such as spin-orbit interaction[5-8], spin Hall effect of light[9-15], topological photonics[16-20], etc. Therein, the transverse spin[21-26], which is studied extensively recently and distinguishes from the longitudinal spin in free space elliptically polarized light beams, plays a significant role in the spin-orbit interaction for a confined electromagnetic field. The transverse spin in an evanescent wave exhibits universal spin-momentum locking and unidirectional excitation known as intrinsic quantum spin Hall effect of light[19, 27]. These features bring about a great deal of applications, such as optical manipulation, metrology, chiral detection, unidirectional transportation, etc.[28-34].

The technique for characterizing the transverse spin density in the focal plane (transverse direction notated as in-plane with respect to the propagating direction $z$) of paraxial beams and tightly-focused beams with gold/silicon nanospheres was reported thereafter[35, 36]. However, it was recently demonstrated that the SAM becomes more complex when the confined field is with a vortex phase thus carrying OAM. For example, the spin-orbit interaction will form many intriguing optical spin structures such as photonic Skyrmions[20]. Indeed, it was demonstrated that the SAM of a structured surface plasmon polariton (SPP) field is proportional to the *curl* of Poynting vector and this spin vector can be regarded as transverse spin universally[20, 37]. Thus, there is always accompanying with the out-of-plane component of SAM in a confined field in the presence of OAM. This requires a new technique to characterize this out-of-plane spin density in order to uncover the full spin vector and the underlying spin-orbit interactions.

In this paper, we proposed a near-field imaging technique to map the out-of-plane spin density of SPPs[38]. The out-of-plane SAM has an inherent relationship with the two circular polarization components of the in-plane field, which lays the foundation for the characterization of SAM. A nanoparticle-on-film structure was employed as a near-field probe, to specifically be sensitive to the in-plane electric field component. The spin-orbit interaction during the excitation of SPPs with various vector vortex beams was uncovered via mapping the SAM. In addition, we find an inherent relationship between the out-of-plane and in-plane SAMs. Thus, the in-plane SAM components can be reconstructed from the measured out-of-plane component and the spin vector structure can be obtained. The proposed method, which can be utilized to reconstruct the topological structures such as optical Skyrmion, is straightforward and effective. It is expected to be valuable for study of spin optics and topological photonics.

## Theoretical basis

SPP is a transverse magnetic (TM) evanescent mode supported at a metal-dielectric interface. Considering that the SPP is propagating in the *xy*-plane and exponentially decaying in the upper half-plane z > 0 (as shown in Fig. 1(a)), the electromagnetic (EM) fields should fulfil the following relationship[39]:

$$E_x = -\frac{k_z}{k_r^2}\frac{\partial E_z}{\partial x} \quad H_x = -\frac{i\omega\varepsilon}{k_r^2}\frac{\partial E_z}{\partial y}$$
$$E_y = -\frac{k_z}{k_r^2}\frac{\partial E_z}{\partial y} \quad H_y = \frac{i\omega\varepsilon}{k_r^2}\frac{\partial E_z}{\partial x}, \quad (1)$$

Note here that we express the EM field in the Cartesian coordinates (*x*, *y*, *z*) with directional unit vector $(\hat{\mathbf{e}}_x, \hat{\mathbf{e}}_y, \hat{\mathbf{e}}_z)$, while the SPPs in other coordinate systems can also be derived with the similar procedure. Here, $E_z$ is the *z*-component (out-of-plane) electric field of the SPP wave and it satisfies the Helmholtz equation: $\nabla^2 E_z + k^2 E_z = 0$. In addition, the out-of-plane magnetic field ($H_z$) of a TM-mode vanishes. Here, $k_r^2 = k^2 + k_z^2$, where $ik_z$, $k_r$ and $k$ are the out-of-plane, the in-plane and the total wave-vector, respectively; $\omega$ stands for the angular frequency of the EM field; $\varepsilon$ is the permittivity of the propagating medium.

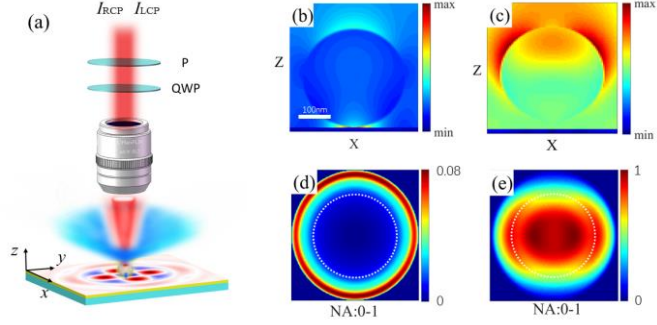

**Fig.1**. (a) Schematic diagram of the optical system for mapping the near-field distributions of SAM. The SPP is excited by a focused vector beam with wavelength of 632.8nm in the air/gold interface. A PS nanosphere is fixed on the metal film, which is controlled by a Piezo scanning stage, to scatter the near-field SPP electric field to the far field. The radiation field of in-plane electric field component is collected by a low NA objective as a low-pass filter. At last, the intensities of the two circularly polarized components ($I_{RCP}$ and $I_{LCP}$) of the radiation field, are respectively filtered out by a quarter wave plate (QWP) and a polarizer (P) and measured by a photo-multiplier tube to obtain the z-component SAM. (b) and (c) The simulated electric field distributions near a polystyrene nanosphere on gold film structure when excited by the out-of-plane and in-plane fields, respectively. The radius of the nanosphere is 160nm and the gap between the nanosphere and gold film is 2nm. (d) and (e) The corresponding far field scattering radiation patterns in the Fourier space. The white dashed circles represent a NA of 0.7, which is employed in the experiment for collecting the scattering signals. QWP: quarter-wave plate, P: polarizer

Obviously, the in-plane electric and magnetic field components have the following relations:

$$H_x = \frac{i\omega\varepsilon}{k_z}E_y, \quad H_y = -\frac{i\omega\varepsilon}{k_z}E_x . \tag{2}$$

The SAM of an arbitrary EM wave can be calculated by $\mathbf{S} = \{\varepsilon \mathbf{E}^* \times \mathbf{E} + \mu \mathbf{H}^* \times \mathbf{H}\}/4\omega$[4], where * stands for the complex conjugate and $\mu$ is permeability of the medium. As the result, the z-component SAM is related to the in-plane circularly polarized (CP) components by:

$$S_z = S_z^e + S_z^h = \frac{\varepsilon}{4\omega i}\frac{k_r^2}{k_z^2}\left(E_x^* E_y - E_y^* E_x\right) = \frac{\varepsilon}{4\omega}\frac{k_r^2}{k_z^2}\left(I_{RCP} - I_{LCP}\right), \tag{3}$$

where $I_{RCP}$ and $I_{LCP}$ indicates the right and left circularly polarized components, respectively. In a word, the out-of-plane SAM of a SPP field can be uncovered by simply measuring the right-handed and left-handed CP components of the in-plane electric field distribution.

Subsequently, by carefully examining the expression of spin vectors, it can be deduced that (see the Supplemental Note 1 for the detail of derivation)

$$\frac{\partial S_x}{\partial y} = \frac{\partial S_y}{\partial x} . \tag{4}$$

By employing the continuity equation of the spin vector: $\nabla \cdot \mathbf{S} = 0$, we can obtain another relationship between the z-component and in-plane SAM components:

$$\frac{\partial S_x}{\partial x} + \frac{\partial S_y}{\partial y} = 2k_z S_z . \tag{5}$$

From Eqs. (4-5), for a TM mode evanescent wave such as SPP, the in-plane and out-of-plane SAM components have a relationship of:

$$\frac{\partial^2 S_x}{\partial x^2} + \frac{\partial^2 S_x}{\partial y^2} = 2k_z \frac{\partial S_z}{\partial x}, \tag{6}$$

$$\frac{\partial^2 S_y}{\partial x^2} + \frac{\partial^2 S_y}{\partial y^2} = 2k_z \frac{\partial S_z}{\partial y}. \tag{7}$$

Thus, once the out-of-plane component ($S_z$) was measured, the in-plane SAM components ($S_x$, $S_y$) can be reconstructed by solving the linear partial differential equations (6-7) (see Supplemental Note 1 for the detail of solving).

## Experimental results and discussion

The schematic diagram of the mapping configuration is given in Fig. 1(a), where a polystyrene (PS) nanosphere with a radius of 160nm was immobilized on a thin gold film to scatter the SPPs excited on the surface (see Supplemental Note 2 for the detail of setup). The SPPs field are excited by various cylindrical vector beams (CVBs) with vortex phase[40] under a focusing configuration in the metal/air interface. The generalized formula of cylindrical vector vortex beams can be expressed in the cylindrical coordinates $(r, \varphi, z)$ with directional unit vector $(\hat{\mathbf{e}}_r, \hat{\mathbf{e}}_\varphi, \hat{\mathbf{e}}_z)$ as:

$$\mathbf{E}_i = A_r e^{im\varphi} \left[ \eta_r \cos\{(n-1)\varphi + \varphi_0\} \hat{\mathbf{e}}_r + \eta_\varphi \sin\{(n-1)\varphi + \varphi_0\} \hat{\mathbf{e}}_\varphi \right], \tag{8}$$

where $A_r$ is the envelope of amplitude and in general has a Bessel-Gauss function envelope. $\eta_r$ and $\eta_\varphi$, which are the weight factors of the radial and azimuthal polarized components, can be real or imaginary numbers. $\varphi_0$ gives the original polarized phase of light beams, $n$ denotes the polarization topological charge (PTC), $m$ indicates the vortex topological charge (VTC) of incident light defined in the cylindrical coordinate system.

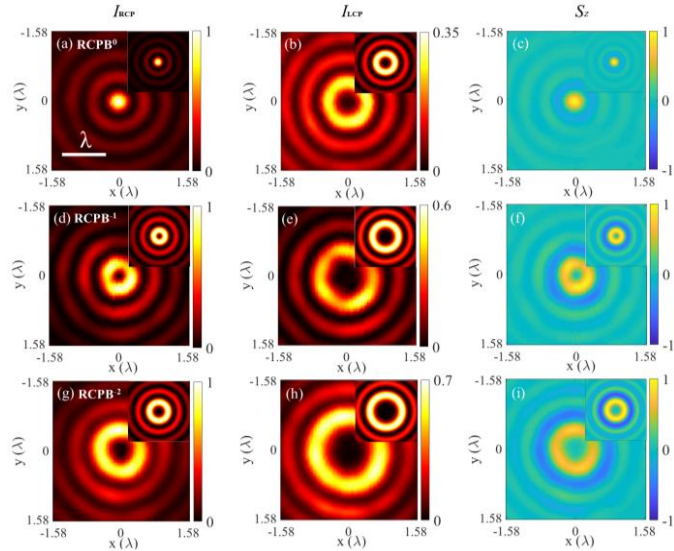

**Fig.2.** The experimental results for the SPPs that are excited by (a-c) right-handed circularly polarized beam without vortex phase (RCPB$^0$), (d-f) with -1-order vortex phase (RCPB$^{-1}$) and (g-i) with -2-order vortex phase (RCPB$^{-2}$), respectively. The left and middle panels show the measured right-handed ($I_{RCP}$) and left-handed ($I_{LCP}$) CP component of the in-plane field of SPPs, and the right panels show the resultant z-component of SAM. The corresponding theoretically calculated distributions are shown in the insets, for the purpose of comparison. The mapping region is 2μm×2μm and the step size is 20nm for all experimental results. The scale is normalized to the wavelength λ as shown in the inset of (a).

Nanoantennas, such as PS and silicon nanosphere, can support various resonances at visible frequencies. In particular, the nanoantenna-on-metal structure is of special interest since its scattering can be shaped by the full vectorial nature of the excited field[41-43]. In this work, a PS nanosphere-on-film structure was employed as a near-field probe. Generally, the out-of-plane electric field component of SPPs would couple to the gap mode between the metal film and nanosphere (Fig. 1(b)), while the in-plane field component will induce an electric resonance within the dielectric particles (Fig. 1(c)). The strongly localized gap mode excited by the out-of-plane field has a higher effective index and hence a higher effective **k**-vector comparing to that of the resonant mode excited by the in-plane field. A larger wave-vector will result in a larger scattering angle in the Fourier space as shown in Figs. 1(d-e). This polarization-dependent directional scattering of the two types of modes provides us an approach to separate the scattering signals that are induced by the in-plane and out-of-plane components of electric field vector. An objective lens with numerical aperture (NA) of 0.7 was employed to collect the radiate waves from the nanosphere. Under this circumstance, most of the scattering radiations induced by the out-of-plane electric field are rejected by the system, with the collected signals merely reflecting the information of the in-plane field component of SPPs. After collected by the objective lens, the scattering radiation from the nanosphere that contains the spin information of SPPs is analyzed by a combination of quarter wave plate and linear polarizer. By changing the angle between the fast axis of the wave plate and the axis of the polarizer, the two CP components can be filtered out respectively, to reveal the local SAM of SPPs at the position of the nanosphere.

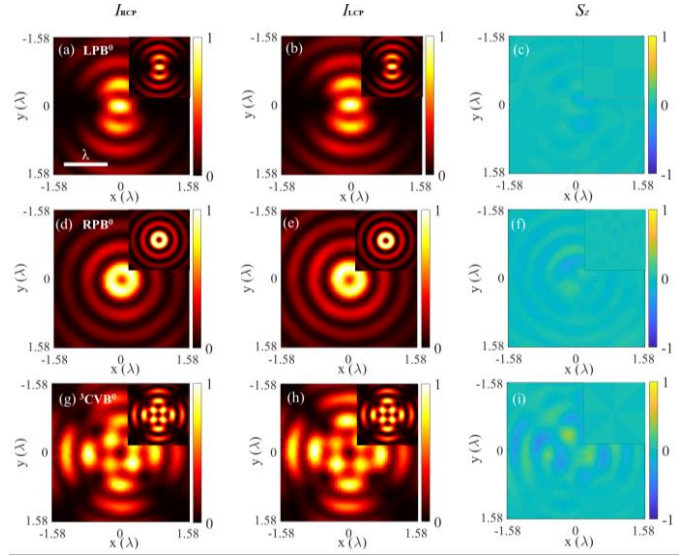

**Fig.3**. The experimental and simulated results for the SPPs that are excited by (a-c) linearly polarized beam (LPB$^0$), (d-f) radially polarized beam (RPB$^0$) and (g-i) 3$^{rd}$-order CVB without vortex ($^3$CVB$^0$), respectively. The left and middle panels show the measured right-handed ($I_{RCP}$) and left-handed ($I_{LCP}$) CP component of the in-plane field of SPPs, and the right panels show the resultant z-component of SAM. The mapping region is 2μm×2μm and the step size is 20nm for all experimental results. The scale is normalized to the wavelength λ as shown in the inset of (a).

Firstly, we consider the measurement of the SPPs excited by a right-handed circularly polarized beam (RCPB$^0$: here and in the following, the right-side superscript indicates the vortex phase of incident light). The experimental mapping results of the two CP components of the in-plane SPP field are shown in Figs. 2(a-b). For comparison, the insets show the theoretical ones calculated with the Richard-Wolf vectorial diffraction theory (see Supplemental Note 3 for the detail of calculation of the focused fields). Note that the intensity of the left-handed CP component is weaker than that of the right-handed one. The out-of-plane SAM of the SPPs can thus be obtained by using Eq. (6), with the result shown in Fig. 2(c). It can be seen that the experimental results match excellently with theory. From Eq. (8), it can be obtained that the RCPB$^0$ is corresponding to $n$=1, $m$=1, $\eta_r$=1, $\eta_\varphi$=$i$ and $\varphi_0$=45°, respectively. Here, the additional VTC ($m$=1) results from the coordinate transform from the Cartesian to the cylindrical coordinate system. Generally, the focusing of RCPB$^0$ can present a right-handed CP component at the focal plane (Fig. 2(a)), and the distribution at the focal plane for this component is solid. More interesting is that the left-handed CP component also appears owing to the Berry phase induced under a focusing process (Fig. 2(b)). This component carries additional vortex phase (see Eq. (S16) in Supplemental Note 4), which causes the profile of the focused field hollow. This is indeed a manifestation of the spin-to-orbital AM conversion.

When the incident RCP beam bears a vortex phase, and thus carries the OAM, this OAM in the incident beam will transfer to the SPP, causing its right-handed CP components with a central hollow distribution. Meanwhile, because the left-handed CP component carries additional vortex phase from the incident beam, the radii of its maximum intensity ring will be enlarged or it will become a central solid spot depends on the handedness of the vortex phase. In Figs. 2(d-f) and Figs. 2(g-i) we show the experimental mapping results when the incident RCP beam carries a −1-order vortex phase (RCPB$^{-1}$) and a −2-order vortex phase (RCPB$^{-2}$), respectively, which verifies the above analysis. It is worthy to be noted that the ratio between the maximal intensities of the left-handed and right-handed CP components increases as the VTC scales up. This indicates that the weight of the reversed spin with respect to the incident one lifts up, manifesting a more remarkable spin-orbit interaction.

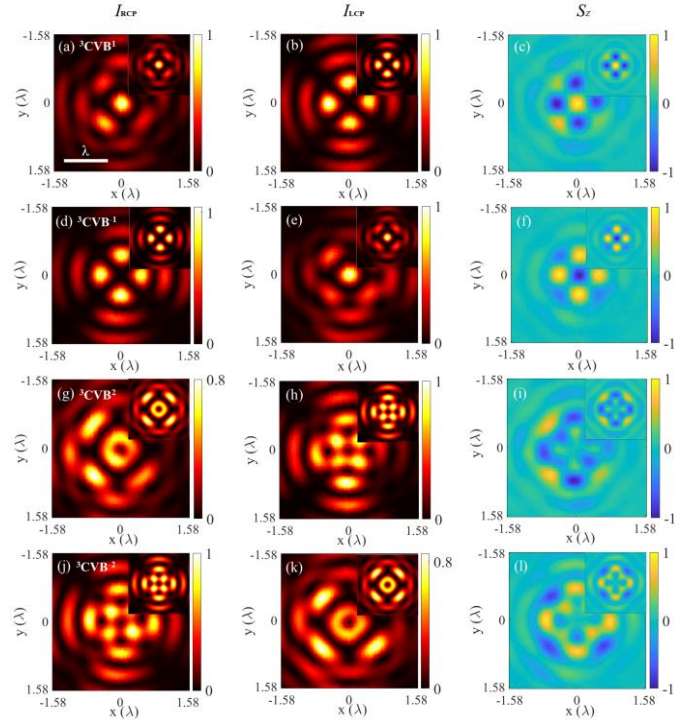

**Fig.4**. The experimental and simulated results for the SPPs that are excited by a 3$^{rd}$-order CVB with (a-c) 1$^{st}$-order vortex phase ($^3$CVB$^1$), (d-f) -1$^{nd}$-order vortex phase ($^3$CVB$^{-1}$), (g-i) 2$^{nd}$-order vortex phase ($^3$CVB$^2$), and (j-l) -2$^{nd}$-order vortex phase ($^3$CVB$^{-2}$), respectively. The left and middle panels show the measured right-handed ($I_{RCP}$) and left-handed ($I_{LCP}$) CP component of the in-plane field of SPPs, and the right panels show the resultant z-component of SAM. The mapping region is 2μm×2μm and the step size is 20nm for all experimental results. The scale is normalized to the wavelength λ as shown in the inset of (a).

Subsequently, we consider the measurement of the out-of-plane SAM of SPPs that are excited by a linearly polarized beam (LPB$^0$: $n=0$, $m=0$, $\eta_r=1$, $\eta_\varphi=1$ and $\varphi_0=90°$), a radially polarized beam (RPB$^0$: $n=1$, $m=0$, $\eta_r=1$, $\eta_\varphi=0$ and $\varphi_0=0°$) and a high-order cylindrical vector beam ($^3$CVB$^0$: $n=3$, $m=0$, $\eta_r=1$, $\eta_\varphi=1$ and $\varphi_0=0°$), respectively. Note here that the left-side superscript gives the PTC of the beams. By examining Eq. (S19) in Supplemental Note 4, it is observed that these three beams can be decomposed into equally-weighted right-handed and left-handed CP components. Under the focusing condition, the two helical components will produce a pair of opposite helices components with equal magnitude. This results in the absence of the out-of-plane SAMs in the excited SPPs. The experimentally-mapped CP components of the SPPs for the incident $LPB^0$, $RPB^0$ and $^3CVB^0$ are shown in Figs. 3(a-b), Figs. 3(d-c), and Figs. 3(g-h), respectively, and the corresponding deduced out-of-plane SAM components are given in Figs. 3(c), 3(f) and 3(i). One can clearly find the same distributions for the two CP components, validating the absence of the out-of-plane SAM in the SPPs.

The situation becomes different when the incident CVBs carry vortex phase. If the VTC is nonzero, the integral of the phase items $e^{i(m\pm 1)\varphi}$ will cause the separation of the two helical components, which makes the out-of-plane component of SAM appear. The experimental mapping results of the focused SPP fields for the $^3CVB^1$, $^3CVB^{-1}$, $^3CVB^2$ and $^3CVB^{-2}$ are shown in Fig. 4, together with the theoretical ones shown in the insets. One can clearly see the different patterns of the two CP components owing to the presence of OAM in the incident beam, which indicates the emergence of out-of-plane SAM in the excited SPPs. Interestingly, the distributions of the right-handed and left-handed CP components exchange when the VTC reversals, resulting in the reversal of the SAM. This is a manifestation of the orbit-to-spin AM conversion.

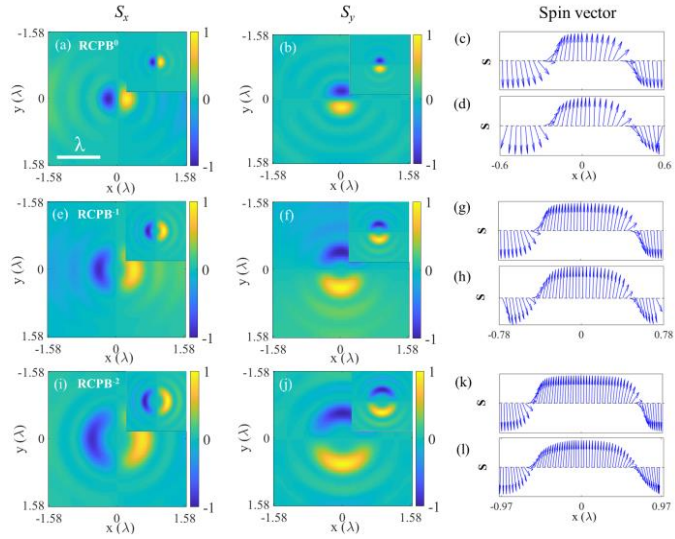

**Fig.5**. The reconstructed in-plane SAM ($S_x$, $S_y$) components of SPPs excited by the (a-b) $RCPB^0$, (e-f) $RCPB^{-1}$ and (i-j) $RCPB^{-2}$, respectively. The left and middle panels show the reconstructed $S_x$ and $S_y$, and the right panels show the theoretical and reconstructed spin vectors for (c-d) $RCPB^0$, (g-h) $RCPB^{-1}$ and (k-l) $RCPB^{-2}$. All three reconstructed spin structure of vector field with different vortex topological charges can also be regarded as photonic Skyrmion and vary from central 'up' states to boundary 'down' states gradually. The calculation region is 2μm×2μm and the step size is 20nm.

Above we have demonstrated that the out-of-plane SAM component of SPPs can be mapped by making use of the inherent spin-polarization relationship. However, in many cases such as in a photonic spin Skyrmion[20], both the in-plane and out-of-plane SAM components need to be characterized in order to get a whole picture of the phenomenon. Unfortunately, it is hard to obtain both of them simultaneously because of the anisotropic response of the probe to the in-plane and out-of-plane fields. An alternative approach is that we find an inherent relationship between the two SAM components and reconstruct the others from the measured one.

The left and middle panels in Fig. 5 show the reconstructed in-plane SAM distributions from the measured out-of-plane SAM in Fig. 2, together with the calculated results shown in the insets. In particular, the directional spin vectors (since the spin vectors of these three fields are rotational symmetric, we only exhibit the spin vector along the x-direction here) definitely show the vortex spin structure analog to the Neel-type Skyrmion. Obviously, all three reconstructed spin structure of vector field with different vortex topological charges can also be regarded as photonic Skyrmion and vary from central 'up' states to boundary 'down' states gradually, which is the first time verified by our work in experiment. Because the intensities of electric field are near to zero in the boundary of photonic Skyrmion, there are tiny experimental mismatches in the boundary. The experimental results demonstrate the good agreement with theoretical ones, which validates the effectiveness of this approach.

## Conclusions

In conclusion, we proposed and demonstrated an approach to map the near field distributions of the optical spin angular momenta. The method is based on the inherent relationship of surface waves and making use of a nanosphere-on-film probe to specifically target on the in-plane field. The experimental results on the SPPs match excellently with the theoretical ones, validating the effectiveness of the proposed method. The spin-orbit interaction during the excitation of SPPs with various vector vortex beams was uncovered via mapping the SAM. We also reconstruct the in-plane SAM components from the measured $z$-component by using the inherent relationship between them. To the best of our knowledge, this is the first demonstration of a method which aims to uncover the out-of-plane SAM and, furthermore, construct the topological spin structure such as photonic Skyrmion of a confined EM field. This will facilitate the exploration of optical spin-orbit interaction at the near field. Furthermore, by fabricating this kind of nanoantenna-on-film structure onto an optical fiber tip to form a scanning nano-probe, it will be allowed to map the SAM of various structures and materials, and together with other near-field imaging technique such as tip-enhanced nano-spectroscopy[44], tip-enhanced SAM spectroscopy can be developed for characterizing the optical chirality, which is expected to be valuable for study of the chiral properties of samples at deep-subwavelength scale.

## Conflicts of interest

There are no conflicts to declare.

## Acknowledgements

This work was partially supported by the National Natural Science Foundation of China (NSFC) (grants Nos. 61490712, 61427819, 61622504, 11504244, 61705135); The leading talents of Guangdong province program (grant 00201505); the Natural Science Foundation of Guangdong Province (grant 2016A030312010); Shenzhen Science and Technology Innovation Commission (grants Nos.KQTD2015071016560101, KQTD2017033011044403, ZDSYS201703031605029, KQTD2018041218324255, JCYJ20180507182035270), Shenzhen University (2019074). L. Du acknowledges the support given by Guangdong Special Support Program.

Supplemental Materials for

# Mapping the near-field spin angular momenta in the structured surface plasmon polaritons field

Congcong Li, Peng Shi, Luping Du* and Xiaocong Yuan*

*Nanophotonics Research Center, Shenzhen Key Laboratory of Micro-Scale Optical Information Technology & Institute of Microscale Optoelectronics, Shenzhen University, Shenzhen 518060, China*

*corresponding author: lpdu@szu.edu.cn and xcyuan@szu.edu.cn

1. **Supplemental Note1 | Relationship between various SAM components**

2. **Supplemental Note2 | Experimental setup and methods**

3. **Supplemental Note3 | Richard-Wolf vectorial diffraction theory in the cylindrical coordinates**

4. **Supplemental Note4 | Richard-Wolf vectorial diffraction theory in the helical coordinates**

**Supplemental Note1 | Relationship between various SAM components**

For transverse magnetic (TM) evanescent modes, using Maxwell's equation, the relation between the transverse electric, magnetic field components and longitudinal electric field component ($E_z$) can be expressed as (from the Eq. 1 of main text):

$$E_x = -\frac{k_z}{k_r^2}\frac{\partial E_z}{\partial x} \quad H_x = -\frac{i\omega\varepsilon}{k_r^2}\frac{\partial E_z}{\partial y}$$
$$E_y = -\frac{k_z}{k_r^2}\frac{\partial E_z}{\partial y} \quad H_y = \frac{i\omega\varepsilon}{k_r^2}\frac{\partial E_z}{\partial x}$$
, (S.1)

where $\omega$ and $\varepsilon$ give the angular frequency and permittivity of surround material. Here, $k_r$ and $ik_z$ denote the transverse and longitudinal components of wave-vector **k** with $k = |\mathbf{k}| = \sqrt{k_r^2 - k_z^2}$. Using the above relationships, the spin angular momentum, which is expressed as

$$\mathbf{S} = \frac{1}{4\omega}\mathrm{Im}\{\varepsilon\mathbf{E}^* \times \mathbf{E} + \mu\mathbf{H}^* \times \mathbf{H}\},$$ (S.2)

can be calculated to be:

$$\begin{cases} S_x = 2\mathrm{K}k_z\,\mathrm{Im}\left(E_z^*\frac{\partial E_z}{\partial y}\right) \\ S_y = 2\mathrm{K}k_z\,\mathrm{Im}\left(E_z\frac{\partial E_z^*}{\partial x}\right), \\ S_z = 2\mathrm{K}\,\mathrm{Im}\left(\frac{\partial E_z^*}{\partial x}\frac{\partial E_z}{\partial y}\right) \end{cases}$$ (S.3)

where $\mathrm{K} = \varepsilon/4\omega k_r^2$ is a physical constant.

By carefully examining the expression of spin vectors, we find that

$$\frac{\partial S_x}{\partial y} = 2\mathrm{K}k_z\,\mathrm{Im}\left(\frac{\partial E_z^*}{\partial y}\frac{\partial E_z}{\partial y} + E_z^*\frac{\partial^2 E_z}{\partial^2 y}\right) = 2\mathrm{K}k_z\,\mathrm{Im}\left(E_z^*\frac{\partial^2 E_z}{\partial^2 y}\right),$$ (S.4a)

and

$$\frac{\partial S_y}{\partial x} = 2\mathrm{K}k_z\,\mathrm{Im}\left(\frac{\partial E_z}{\partial x}\frac{\partial E_z^*}{\partial x} + E_z\frac{\partial^2 E_z^*}{\partial x^2}\right) = 2\mathrm{K}k_z\,\mathrm{Im}\left(E_z\frac{\partial^2 E_z^*}{\partial x^2}\right).$$ (S.4b)

On the other hand, the z-component electric field $E_z$ satisfies Helmholtz equation:

$$\nabla^2 E_z + k^2 E_z = \frac{\partial^2 E_z}{\partial x^2} + \frac{\partial^2 E_z}{\partial y^2} + k_r^2 E_z = 0.$$ (S.5)

Therefore, by employing the Eq. (S5), the Eq. (S4a) can be converted into

$$\frac{\partial S_x}{\partial y} = 2\mathrm{K}k_z\,\mathrm{Im}\left(-E_z^*\frac{\partial^2 E_z}{\partial x^2} - k_r^2 E_z^* E_z\right) = 2\mathrm{K}k_z\,\mathrm{Im}\left(E_z\frac{\partial^2 E_z^*}{\partial x^2}\right) = \frac{\partial S_y}{\partial x}.$$ (S.6)

This is a general relation for the TM-mode surface wave, and even for the TE-mode surface wave, a similar relation can be obtained. Then, by employing the conservation law of the spin vectors ($\nabla \cdot \mathbf{S} = 0$ which can be deduced by equation S.3), we can obtain a pair of linear partial differential equations for transverse spin vectors:

$$\begin{cases} \dfrac{\partial^2 S_x}{\partial x^2} + \dfrac{\partial^2 S_x}{\partial y^2} = 2k_z \dfrac{\partial S_z}{\partial x} \\ \dfrac{\partial^2 S_y}{\partial x^2} + \dfrac{\partial^2 S_y}{\partial y^2} = 2k_z \dfrac{\partial S_z}{\partial y} \end{cases}. \quad (S.7)$$

Note here that the $S_z$ can be mapped experimentally. Thus, the transverse SAM $S_x$ and $S_y$ can be obtained by solving the linear partial differential equation pair. The eigenfunction set of the equation (S.6) can be obtained by the separation of variables method.

Then the transverse SAM $S_x$ can be expressed by a series of Fourier expansions:

$$S_x = \sum_{n=-\infty}^{\infty} \sum_{m=-\infty}^{\infty} \left( A_m \sin \frac{m\pi x}{L_x} + B_m \cos \frac{m\pi x}{L_x} \right) \left( C_n \sin \frac{n\pi y}{L_y} + D_n \cos \frac{n\pi y}{L_y} \right), \quad (S.8)$$

where $L_x$ and $L_y$ are boundary condition parameters, which are equal to the measured region $2\mu m \times 2\mu m$ in the work. $A_m$, $B_m$, $C_m$ and $D_m$ are undetermined constants.

From the first formula of equation (S.7) and equation (S.8), we can get that:

$$-\sum_{n=-\infty}^{\infty} \sum_{m=-\infty}^{\infty} \lambda_{mn} \left( A_m \sin \frac{m\pi x}{L_x} + B_m \cos \frac{m\pi x}{L_x} \right) \left( C_n \sin \frac{n\pi y}{L_y} + D_n \cos \frac{n\pi y}{L_y} \right) = 2k_z \frac{\partial S_z}{\partial x}, \quad (S.9)$$

where $\lambda_{mn} = (\frac{m\pi}{L_x})^2 + (\frac{n\pi}{L_y})^2$ is the eigenvalue.

Actually, $S_z$ is mirror symmetric with respect to $x$, $y$-axes, which can be expressed as $\hat{m}_x S_z = S_z$ and $\hat{m}_y S_z = S_z$, where $\hat{m}_i$ is the mirror operator with respect to $i$-axis in Cartesian coordinates. It need be noted that the $i$-coordinate suffers mirror antisymmetric with respect to $i$-axis, while it suffers mirror symmetric with respect to another axis, so that we have $\hat{m}_x x = -x$ and $\hat{m}_y x = x$. Thus we can get the following relationships:

$$\begin{aligned} \hat{m}_x \frac{\partial S_z}{\partial x} &= -\frac{\partial S_z}{\partial x} \\ \hat{m}_y \frac{\partial S_z}{\partial x} &= \frac{\partial S_z}{\partial x} \end{aligned}, \quad (S.10)$$

which can be utilized to simplify the equation (S.9).

Then, equation (S.9) can be modulated to be:

$$-\sum_{n=-\infty}^{\infty}\sum_{m=-\infty}^{\infty}\lambda_{mn}\left(A_m \sin\frac{m\pi x}{L_x}\cos\frac{n\pi y}{L_y}\right) = 2k_z\frac{\partial S_z}{\partial x} \quad . \tag{S.11}$$

Up to now, the expansion coefficient $A_m$ can be determined through the Fourier integral of equation (S.11). Omitting this complex mathematical process, we obtain that the $S_x$ can be solved to be:

$$\begin{cases} S_x = \sum_{n=-\infty}^{\infty}\sum_{m=-\infty}^{\infty}\left(A_m \sin\frac{m\pi x}{L_x}\cos\frac{n\pi y}{L_y}\right) \\ A_m = -\frac{1}{4L_xL_y\lambda_{mn}}\int_0^{L_x}\int_0^{L_y} 2k_z\frac{\partial S_z}{\partial x}\sin\frac{m\pi x}{L_x}\cos\frac{n\pi y}{L_y}dxdy \end{cases} . \tag{S.12}$$

In a similar manner, the $S_y$ can be expressed as:

$$\begin{cases} S_y = \sum_{n=-\infty}^{\infty}\sum_{m=-\infty}^{\infty}\left(B_m \cos\frac{m\pi x}{L_x}\sin\frac{n\pi y}{L_y}\right) \\ B_m = -\frac{1}{4L_xL_y\lambda_{mn}}\int_0^{L_x}\int_0^{L_y} 2k_z\frac{\partial S_z}{\partial y}\cos\frac{m\pi x}{L_x}\sin\frac{n\pi y}{L_y}dxdy \end{cases} . \tag{S.13}$$

Through the measured longitudinal SAM, the transverse SAM can be reconstructed using equation (S.12) and equation (S.13), and the photonic Skyrmion can be recognized. For numerical calculation, we assume that the parameters $L_x$ and $L_y$ is $2\mu m$, considering the intensity of the SAM at this position is negligible.

**Supplemental Note2 | Experimental setup and methods**

The experiment is carried out by a laser optical source with wavelength equal to 632.8nm. After a telescope system, multiple half-wave plates (HWP) or quarter-wave plates (QWP) and vortex wave plates (VWP) are employed to generate the various vector beams. A reflecting liquid crystal spatial light modulator (SLM) is utilized to add desired vortex phase to the vector beams. Then, the vector beams with phase are tightly focused by an oil-immersion objective (Olympus, NA=1.49, 100×) to excite the SPP (surface plasmon polariton) waves at the air/metal interface. The PS nanosphere-on-film is prepared and fixed on a Piezo scanning stage (Physik Instrumente, P-545) to scatter the near-field SPP electric field to the far field and collected the transverse component by an objective (Olympus, NA=0.7, 60×) as a low-pass filter. At last, the intensities of the two circularly polarized components ($I_{RCP}$ and $I_{LCP}$) of the radiation field, are respectively filtered out by a quarter wave plate (QWP) and a polarizer (P) and measured by a photo-multiplier tube (PMT, Hamamatsu R12829). By raster scanning the piezo-stage, the $I_{RCP}$ and $I_{LCP}$ can respectively be mapped to reconstruct the near-field distribution of the longitudinal SAM component.

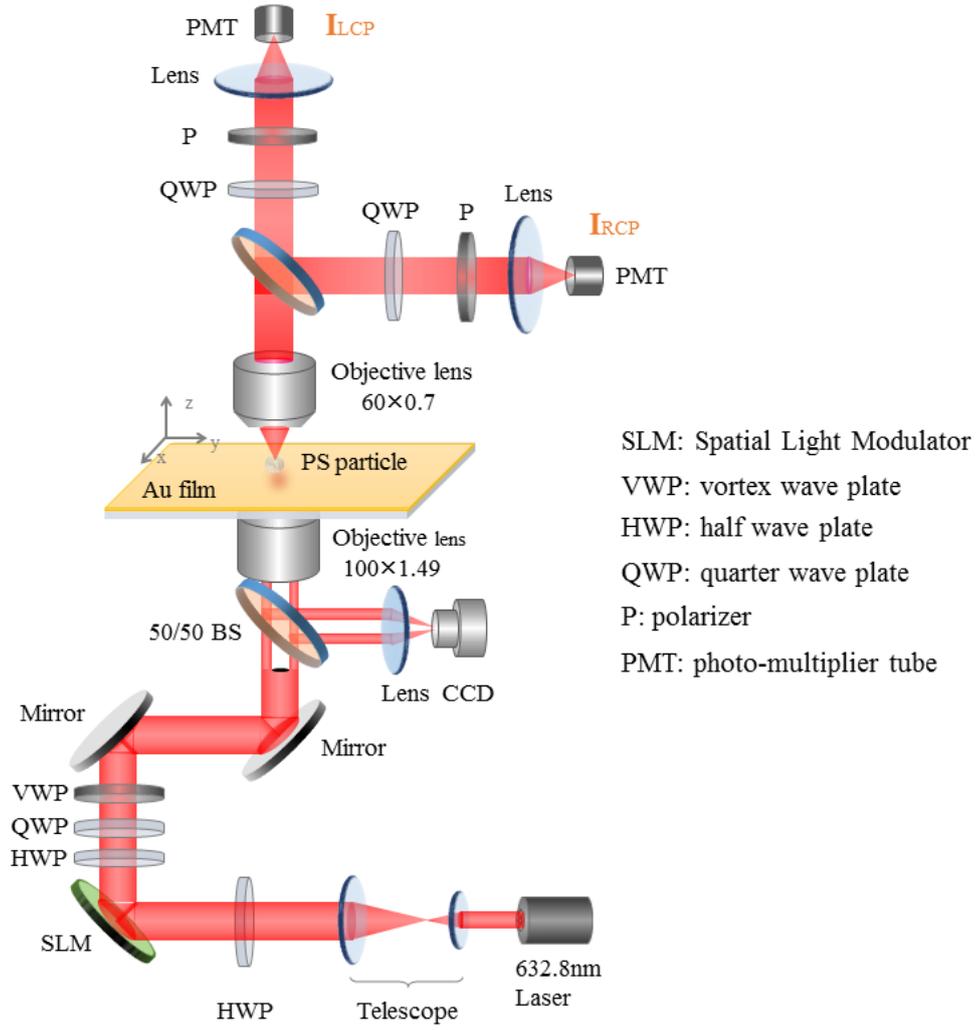

**Fig. S1.** The set-up for mapping the longitudinal SAM component.

## Supplemental Note3 | Richard-Wolf vectorial diffraction theory in the cylindrical coordinates

Here, we utilize the Richard-Wolf vectorial diffraction theory to calculate the focused SPP field[1]. For an arbitrary incident light given by

$$\mathbf{E}_i = A_r \begin{bmatrix} \eta_r \cos\{(n-1)\varphi+\varphi_0\}\hat{\mathbf{e}}_r \\ +\eta_\varphi \sin\{(n-1)\varphi+\varphi_0\}\hat{\mathbf{e}}_\varphi \end{bmatrix} e^{im\varphi}. \quad (S.14)$$

We calculate the field and polarization distributions of the linearly y-polarized beam ($n=0$, $m=0$, $\varphi_0=\pi/2$, $\eta_r=1$, $\eta_\varphi=1$: labelled as LPB$^0$), radially polarized beam ($n=1$, $m=0$, $\varphi_0=0$, $\eta_r=1$, $\eta_\varphi=1$: labelled as RPB$^0$) and 3-order CVB ($n=3$, $m=0$, $\varphi_0=0$, $\eta_r=1$, $\eta_\varphi=1$: labelled as $^3$CVB$^0$) here. The other beams discussed in the manuscript can be superposed by the former beams with an additional vortex phase. The incident fields are exhibited in the Fig. S2.

Subsequently, the spectrum after the lens can be expressed as [2]

$$\mathbf{E}_\infty = A_r e^{im\varphi} \left[ \begin{array}{l} t^p(\theta)\eta_r \cos\{(n-1)\varphi+\varphi_0\} \begin{pmatrix} \cos\varphi\cos\theta \\ \sin\varphi\cos\theta \\ -\sin\theta \end{pmatrix} \\ +t^s(\theta)\eta_\varphi \sin\{(n-1)\varphi+\varphi_0\} \begin{pmatrix} -\sin\varphi \\ \cos\varphi \\ 0 \end{pmatrix} \end{array} \right] \sqrt{\frac{n_i \cos\theta}{n_o}}, \quad (S.15)$$

where $\theta$ and $\varphi$ (in accord with the azimuthal coordinate of incident field) are the orientation angle and the azimuth angle of the focused spherical coordinates, $t^p(\theta)$ and $t^s(\theta)$ indicate the Fresnel transmission coefficients of multilayered film focusing configuration for the *p*-polarization and *s*-polarization light, respectively. In our manuscript, only the *p*-polarization light can pass the insulator-metal-insular structure and excite the SPP field. Thus, $t^s(\theta)=0$. $n_i$ and $n_o$ represent the refractive indices of incident space and focused space, respectively. Then, from the Richard-Wolf vectorial diffraction theory, the focused field can be calculated to be

$$\mathbf{E}_f(\rho,\phi,z) = -\frac{ikfe^{-ikf}}{2\pi} \int_0^{\theta_{max}} \int_0^{2\pi} \mathbf{E}_\infty(\theta,\varphi) e^{ikz\cos\theta} e^{ik\rho\sin\theta\cos(\varphi-\phi)} \sin\theta \, d\varphi \, d\theta, \quad (S.16)$$

where $\theta_{max}$ determined by the numerical aperture (NA) can be expressed as $\theta_{max}=\text{asin}(NA/n_i)$. By performing the two-dimensional integral, one can obtain the focused SPP field precisely.

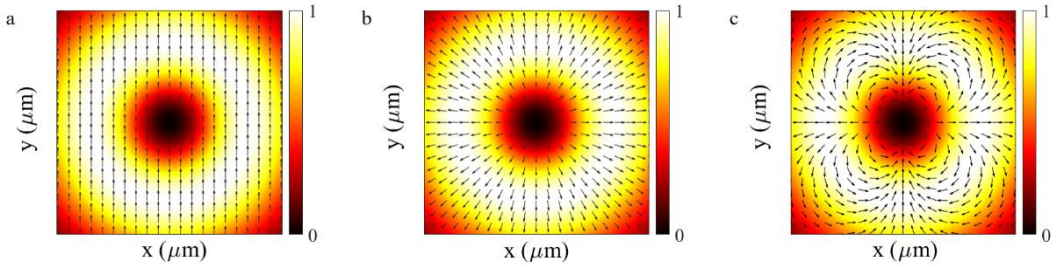

**Fig. S2.** The total field and polarization distributions of (a) linearly y-polarized beam, (b) radially polarized beam and (c) three-order cylindrical vector beam. The field distribution is normalized to the maximal intensity.

**Supplemental Note4 | Richard-Wolf vectorial diffraction theory in the helical coordinates**

The SPPs field are excited by various cylindrical vector beams (CVBs) with vortex phase under a focusing configuration in the metal/air interface. CVBs with spatially variant polarization states and vortex phase have attracted great interest in the past decade because of their fascinating focusing properties and novel applications[2-4]. The generalized formula of CVBs with vortex phase can be expressed in the cylindrical coordinates ($r$, $\varphi$, $z$) with directional unit vector $(\hat{\mathbf{e}}_r, \hat{\mathbf{e}}_\varphi, \hat{\mathbf{e}}_z)$ as:

$$\mathbf{E}_i = A_r e^{im\varphi} \begin{bmatrix} \eta_r \cos\{(n-1)\varphi + \varphi_0\} \hat{\mathbf{e}}_r \\ \eta_\varphi \sin\{(n-1)\varphi + \varphi_0\} \hat{\mathbf{e}}_\varphi \end{bmatrix}, \quad (S.17)$$

where $A_r$ is the envelope of amplitude and in general has a Bessel-Gauss function envelope. $\eta_r$ and $\eta_\varphi$, which are the weight factors of the radial and azimuthal polarized components, can be real or imaginary numbers. $\varphi_0$ gives the original polarized phase of light beams, $n$ denotes the polarization topological charge (PTC), $m$ indicates the vortex topological charge (VTC). The fields in the cylindrical coordinates can be translated into those in the intrinsic helix basis $\hat{\boldsymbol{\sigma}}^\pm = (\hat{\mathbf{e}}_x \pm i\hat{\mathbf{e}}_y)/\sqrt{2}$ via the unitary transformation as shown below:

$$\mathbf{E}_i^\sigma = \frac{A_r e^{im\varphi}}{\sqrt{2}} \begin{bmatrix} E_i^+ \hat{\boldsymbol{\sigma}}^+ \\ E_i^- \hat{\boldsymbol{\sigma}}^- \end{bmatrix}. \quad (S.18)$$

where $E_i^\pm = (\eta_r \cos\{(n-1)\varphi + \varphi_0\} \pm \eta_\varphi \sin\{(n-1)\varphi + \varphi_0\}) e^{\pm i\varphi}$ are the $\hat{\boldsymbol{\sigma}}^+$ and $\hat{\boldsymbol{\sigma}}^-$ components of the incident field.

Considering a focusing configuration by a high aperture (NA) objective lens and according to the vectorial diffraction method[5], the far-field spectrum in the vicinity of the focus can be expressed as:

$$\begin{aligned} \tilde{\mathbf{E}}_f^\sigma &= \sqrt{\frac{n_i \cos\theta}{n_o}} \hat{\mathbf{U}}(\theta, \varphi) \tilde{\mathbf{E}}_i^\sigma \\ &= A_r e^{im\varphi} \sqrt{\frac{n_i \cos\theta}{2n_o}} \left( \begin{bmatrix} \cos^2\frac{\theta}{2} \\ -\sin^2\frac{\theta}{2} e^{-2i\varphi} \\ \frac{\sin\theta}{\sqrt{2}} e^{-i\varphi} \end{bmatrix} E_i'^+ + \begin{bmatrix} -\sin^2\frac{\theta}{2} e^{2i\varphi} \\ \cos^2\frac{\theta}{2} \\ \frac{\sin\theta}{\sqrt{2}} e^{i\varphi} \end{bmatrix} E_i'^- \right), \end{aligned} \quad (S.19)$$

where $E_i'^\pm = (t^p \eta_r \cos\{(n-1)\varphi + \varphi_0\} \pm t^s \eta_\varphi \sin\{(n-1)\varphi + \varphi_0\}) e^{\pm i\varphi}$ are the renewed $\hat{\boldsymbol{\sigma}}^+$ and $\hat{\boldsymbol{\sigma}}^-$ components of the incident field in the metal multilayers structure. The first and second items in Eq. (S.19) are derived from the contributions of the $\hat{\boldsymbol{\sigma}}^+$ and $\hat{\boldsymbol{\sigma}}^-$ components of the incident light, respectively. Note here that the conversion between the spin and orbital angular momentum can be observed intuitively from Eq. (S.19). It can find that the focusing of an incident polarized component $E^\sigma$ can generate the $E_z$ component with the charge-$-\sigma$ vortex $e^{-i\sigma\varphi}$ and the oppositely-polarized component $E^{-\sigma}$ with the charge-$-2\sigma$ vortex $e^{-2i\sigma\varphi}$. These are in accord with the spin-to-orbital angular momentum conversion and the property of angular momentum conservation. Furthermore, the spin-to-orbital angular momentum conversion and the appearance of oppositely-polarized component $E^{-\sigma}$ are originated from the spin-redirection Berry phase term[6].